\RequirePackage{lineno}
\documentclass[aps,prd,twocolumn,showpacs,superscriptaddress,groupedaddress]{revtex4}  
\usepackage{graphicx}  
\usepackage{dcolumn}   
\usepackage{bm}        
\usepackage{amssymb}   
\usepackage{amsmath}
\usepackage{comment}
\usepackage{array}
\usepackage{color}
\usepackage[titletoc]{appendix}
\newcommand{\PreserveBackslash}[1]{\let\temp=\\#1\let\\=\temp}
\newcolumntype{C}[1]{>{\PreserveBackslash\centering}p{#1}}
\newcolumntype{R}[1]{>{\PreserveBackslash\raggedleft}p{#1}}
\newcolumntype{L}[1]{>{\PreserveBackslash\raggedright}p{#1}}

\def \ppp {\pi^+\pi^-\pi^0}
\def \jp {J/\psi}

\def \ff {\phi\phi}

\def \gg {\gamma\gamma}
\def \of {\omega\phi}
\def \ee {e^+e^-}

\def \kk {K^+K^-}
\def \pp {\pi^+\pi^-}

\newcommand{\etac}{\eta_{c}}

\newcommand{\psip}{\psi^{\prime}}

\begin{document}

\title{\boldmath Improved measurements of branching fractions for
  $\etac\to\ff$ and $\of$}

\author{
M.~Ablikim$^{1}$, M.~N.~Achasov$^{9,e}$, X.~C.~Ai$^{1}$, O.~Albayrak$^{5}$, M.~Albrecht$^{4}$, D.~J.~Ambrose$^{44}$, A.~Amoroso$^{49A,49C}$, F.~F.~An$^{1}$, Q.~An$^{46,a}$, J.~Z.~Bai$^{1}$
% ...
\\
\vspace{0.2cm}
(BESIII Collaboration)\\
\vspace{0.2cm} {\it
$^{1}$ Institute of High Energy Physics, Beijing 100049, People's Republic of China\\
% ...
$^{4}$ Bochum Ruhr-University, D-44780 Bochum, Germany\\
$^{5}$ Carnegie Mellon University, Pittsburgh, Pennsylvania 15213, USA\\
% ...
$^{9}$ G.I. Budker Institute of Nuclear Physics SB RAS (BINP), Novosibirsk 630090, Russia\\
% ...
$^{44}$ University of Rochester, Rochester, New York 14627, USA\\
% ...
$^{46}$ University of Science and Technology of China, Hefei 230026, People's Republic of China\\
% ...
$^{49}$ (A)University of Turin, I-10125, Turin, Italy; (B)University of Eastern Piedmont, I-15121, Alessandria, Italy; (C)INFN, I-10125 Turin, Italy\\
% ...
$^{a}$ Also at State Key Laboratory of Particle Detection and Electronics, Beijing 100049, Hefei 230026, People's Republic of China\\
% ...
$^{e}$ Also at the Novosibirsk State University, Novosibirsk 630090, Russia.\\
% ...
}
}
\date{\today}

\begin{abstract}
  Using $(223.7\pm1.4)\times10^6$ $\jp$ events accumulated with the
  BESIII detector, we study $\etac$ decays to $\ff$ and $\of$ final
  states. The branching fraction of $\etac\to\ff$ is measured to be
  $Br(\etac\to\ff)=(2.5\pm0.3^{+0.3}_{-0.7}\pm0.6)\times10^{-3}$, where the
  first uncertainty is statistical, the second is systematic, and the
  third is from the uncertainty of $Br(\jp\to\gamma\etac)$. No
  significant signal for the double Okubo-Zweig-Iizuka suppressed decay of $\etac\to\of$ is observed, and the upper limit on the branching
  fraction is determined to be $Br(\etac\to\of)<2.5\times 10^{-4}$ at
  the 90\% confidence level.
\end{abstract}
\maketitle

\section{Introduction}
Our knowledge of the $\etac$ properties is still relatively poor,
although it has been established for more than thirty
years~\cite{firstetac}. Until now, the exclusively measured decays
only sum up to about 63\% of its total decay width~\cite{pdg}.  The
branching fraction of $\etac\to\ff$ was measured for the first time by
the MarkIII collaboration~\cite{mark3}, and improved measurements were
performed at BESII~\cite{bes1etac2pp,bes2etac2pp} with a precision of
about 40\%. The decay $\etac\to\omega\phi$, which is a doubly
Okubo-Zweig-Iizuka (OZI) suppressed process, has not been observed
yet.

Decays of $\etac$ into vector meson pairs have stood as a bewildering
puzzle in charmonium physics for a long time. This kind of decay is
highly suppressed at leading order in Quantum Chromodynamics (QCD), due to the helicity
selection rule (HSR)~\cite{hsr}. Under HSR, the branching fraction for
$\etac\to\ff$ was calculated to be $\sim2\times10^{-7}$
\cite{charmoniumBr}.  To avoid the manifestation of HSR in charmonium
decays, a HSR evasion scenario was proposed~\cite{evasion}. Improved
calculations with next-to-leading order~\cite{etacqcd} and
relativistic corrections in QCD yield branching fractions varying from
$10^{-5}$~\cite{pqcd} to $10^{-4}$~\cite{hightwist}.  Some
nonperturbative mechanisms, such as the light quark mass corrections
\cite{quarkmasscorrection}, the $^3P_0$ quark pair creation mechanism
\cite{tpz} and long-distance intermediate meson loop effects
\cite{mesonloop}, have also been phenomenologically investigated.

However, the measured branching fraction,
$Br(\etac\to\ff)=(1.76\pm0.20)\times10^{-3}$~\cite{pdg,liuzq}, is much
larger than those of theoretical predictions. To help understand the $\etac$
decay mechanism, high precision measurements of the branching fraction
are desirable. In this paper, we present an improved measurement of the
branching fraction of $\etac\to\ff$, and a search for the doubly
OZI suppressed decay $\etac\to\omega\phi$. The analyses are performed
based on $(223.7\pm1.4)\times10^6$ $\jp$ events~\cite{jsinumber}
collected with the BESIII detector.

\section{DETECTOR and MONTE CARLO SIMULATION}
The BESIII experiment at BEPCII~\cite{NIM1} is an upgrade of
BESII/BEPC~\cite{besii}. The detector is designed to study physics in
the $\tau$-charm energy region~\cite{besphysics}. The cylindrical BESIII
detector is composed of a helium gas-based main drift chamber (MDC), a
time-of-flight (TOF) system, a CsI (Tl) electromagnetic calorimeter
(EMC) and a resistive-plate-chamber-based muon identifier with a
superconducting magnet that provides a 1.0 T magnetic field. The
nominal geometrical acceptance of the detector is 93\% of $4\pi$ solid
angle.  The MDC measures the momentum of charged particles with a
resolution of 0.5\% at 1 GeV/c, and provides energy loss (d$E$/dx)
measurements with a resolution better than 6\% for electrons from
Bhabha scattering. The EMC detects photons with a resolution of 2.5\%
(5\%) at an energy of 1 GeV in the barrel (end cap) region.

To optimize event selection criteria and to understand backgrounds, a
{\sc geant4}-based~\cite{boost} Monte Carlo (MC) simulation package,
BOOST, which includes the description of the geometries and material as
well as the BESIII detection components, is used to generate MC
samples. An inclusive $\jp$-decay MC sample is generated to study the
potential backgrounds. The production of the $\jp$ resonance is
simulated with the MC event generator {\sc kkmc}~\cite{kkmc}, while
$\jp$ decays are simulated with {\sc besevtgen}~\cite{evtgen} for
known decay modes by setting the branching fractions to the world
average values~\cite{pdg}, and with {\sc lundcharm}~\cite{lundcharm}
for the remaining unknown decays. The analysis is performed in the
framework of the BESIII offline software system~\cite{boss}, which
handles the detector calibration, event reconstruction and data
storage.

\section{Event selection}
The $\eta_c$ candidates studied in this analysis are produced by $\jp$ radiative
transitions. We search for $\etac\to\ff$ and $\of$ from
the decays $\jp\to\gamma \ff$ and $\gamma\of$, with final states of
$\gamma2(\kk)$ and $3\gamma\kk\pp$, respectively. The candidate events
are required to have four charged tracks with a net charge of 0,
and at least one or three photons, respectively.

Charged tracks in the polar angle region $|\cos\theta|<0.93$ are
reconstructed from the MDC hits. They must have the point of closest
approach to the interaction point within $\pm10$ cm along the beam
direction and 1 cm in the plane perpendicular to the beam
direction. For the particle identification (PID), the ionization
energy deposited ($dE/dx$) in the MDC and the TOF information are
combined to determine confidence levels (C.L.) for the pion and kaon
hypotheses, and each track is assigned to the particle type with the
highest PID C.L. For the decay $\jp\to\gamma\of\to3\gamma\kk\pp$, two
identified kaons are required within the momentum range of 0.3$-$0.9 GeV with an average efficiency of about 8\%. For the decay
$\jp\to\gamma\ff\to\gamma2(\kk)$, no PID is required. The
intermediate states, $\phi$ and $\omega$, are selected using invariant
mass requirements.

Photon candidates are reconstructed by clustering energy deposits in the EMC
crystals. The energy deposited in the nearby TOF counters is included
to improve the photon reconstruction efficiency and energy
resolution. The photon candidates are required to be in the barrel
region ($|\cos\theta|<0.8$) of the EMC with at least 25 MeV total
energy deposition, or in the end cap regions
$(0.86<|\cos\theta|<0.92)$ with at least 50 MeV total energy
deposition, where $\theta$ is the polar angle of the photon. The
photon candidates are, furthermore, required to be separated from all
charged tracks by an angle larger than $10^\circ$ to suppress photons
radiated from charged particles. The photons in the regions
between the barrel and end caps are poorly measured and, therefore,
excluded. Timing information from the EMC is used to suppress
electronic noise and showers that are unrelated to the event.

Kinematic fits, constrained by the total $\ee$ beam energy momentum,
are performed under the $J/\psi\to\gamma2(\kk)$ and $3\gamma\kk\pp$
hypotheses. Fits are done with all photon combinations together with
the four charged tracks. Only the combination with the smallest
kinematic fit $\chi^2_{4C}$ is retained for further analysis, and
$\chi^2_{4C}<100$ (40) for $J/\psi\to\gamma2(\kk)$ ($3\gamma\kk\pp$)
is required. These requirements are determined from MC simulations by
optimizing $S/\sqrt{S+B}$, where $S$ and $B$ are the numbers of
signal and background events, respectively.

\begin{figure*}[!t]
\centering
\includegraphics[width=7.7cm]{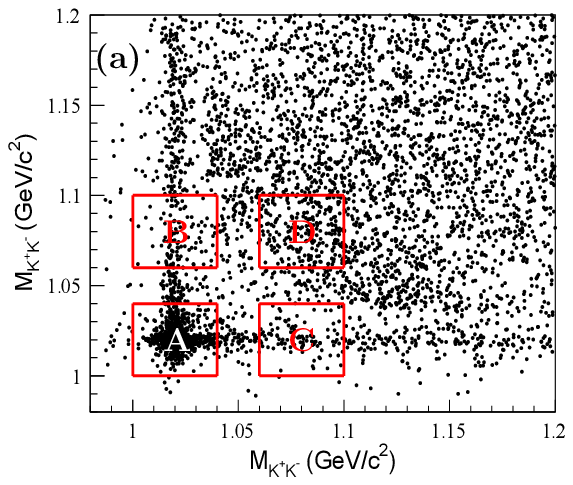}
\hspace{0.2cm}
\includegraphics[height=7cm,width=8.2cm]{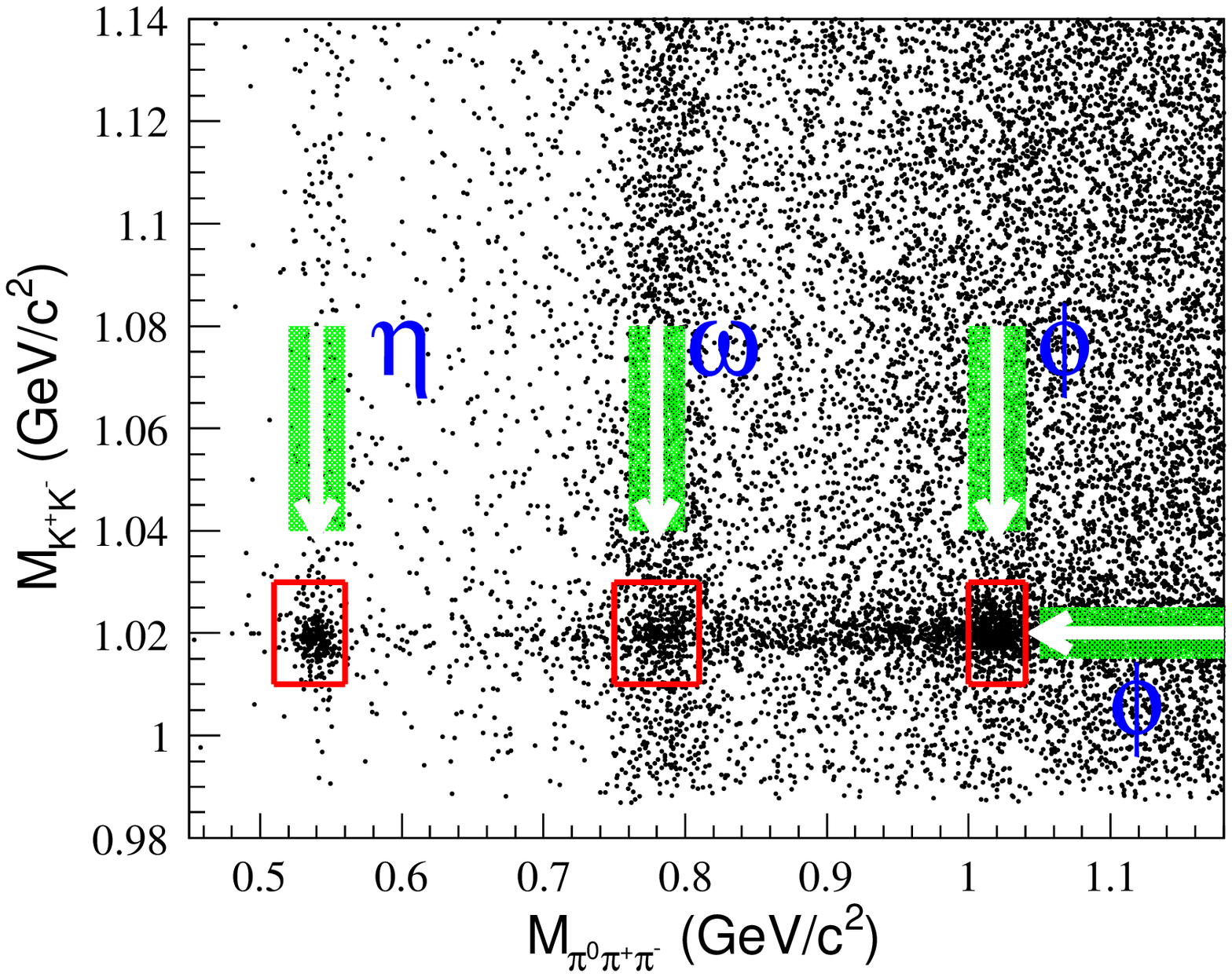}
\put(-185,160){\bf \large (b)} \renewcommand{\figurename}{Fig.}
{\caption{Scatter plot of (a) $M_{K^+K^-}^{(1)}$
    versus $M_{K^+K^-}^{(2)}$ for the decay $\jp\to\gamma2(\kk)$, and (b)
    $M_{K^+K^-}$ versus $M_{\ppp}$ for the decay
    $\jp\to3\gamma\kk\pp$.}
\label{fig:scatterplot}
}
\end{figure*}

Two $\phi$ candidates in the $\jp\to\gamma\ff$ decay are reconstructed
from the selected $2(\kk)$ tracks. Only the combination with a minimum
of $|M_{\kk}^{(1)}-M_{\phi}|^2+|M_{\kk}^{(2)}-M_{\phi}|^2$ is
retained, where $M^{(i)}_{\kk}$ ($i=1,2$) and $M_{\phi}$ denote the
invariant mass of the $\kk$ pair and the nominal mass of the
$\phi$-meson, respectively. A scatter plot of $M_{\kk}^{(1)}$ versus
$M_{\kk}^{(2)}$ for the surviving events is shown in
Fig.~\ref{fig:scatterplot} (a). There is a cluster of events in the
$\ff$ region indicated as a box in Fig. \ref{fig:scatterplot} (a)
originating from the decay $\jp\to\gamma\ff$.
Two $\phi$ candidates are selected by
requiring $|M_{\kk}-M_\phi|<0.02$ GeV/$c^2$, which is determined by
optimizing $S/\sqrt{S+B}$, also.

For the decay $\jp\to\gamma\of\to\gamma\kk\pp\pi^0$, the photon
combination with mass closest to the $\pi^0$ nominal mass is chosen, and
$|M_{\gg}-M_{\pi^0}|<0.02$ GeV/$c^2$ is required. A scatter plot of
the $M_{\kk}$ versus $M_{\ppp}$ for the surviving events is shown in
Fig. \ref{fig:scatterplot} (b). Three vertical bands, as indicated in
the plot, correspond to the $\eta,~\omega$ and $\phi$ decays into
$\ppp$, and the horizontal band corresponds to the decay
$\phi\to\kk$. For the selection of $\jp\to\gamma\of$ candidates, the
$\phi$ and $\omega$ requirements are determined, by optimizing $S/
\sqrt{S+B}$, to be $|M_{\ppp}-M_\omega|<0.03 $ GeV/$c^2$ and
$|M_{\kk}-M_\phi|<0.008$ GeV/$c^2$.

\section{Data analysis}
\subsection{Observation of $\etac\to\ff$}
\begin{figure}[!t]
   \centering
  \includegraphics[width=0.9\columnwidth]{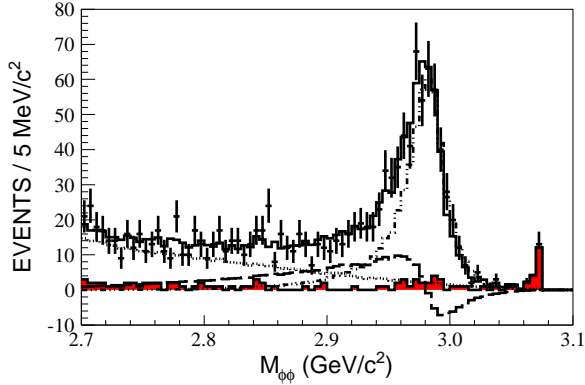}
   \renewcommand{\figurename}{Fig.}
   \caption{Projection of fit results onto the $M_{\ff}$
     spectrum. The dots with error bars denote the data, the solid
     line histogram is the overall result, the dot-dashed histogram is
     the $\eta_c$ signal, the filled red histogram is the combined
     backgrounds estimated with exclusive MC simulations, the dotted
     histogram denotes non $\eta_c$ decays, and the long-dash histogram is
     the interference between the $\etac$ and non $\eta_c$ decays. }
   \label{fig:ffFitResult}
 \end{figure}

Figure~\ref{fig:ffFitResult} shows the invariant mass distribution of
the $\ff$-system within the range from 2.7 to 3.1 GeV/$c^2$. The
$\etac$ signal is clearly observed. Background events from $\jp$
decays are studied using the inclusive MC sample. The dominant
backgrounds are from the decays $\jp\to\gamma\phi\kk$ and
$\jp\to\gamma\kk\kk$ with or without an $\eta_c$ intermediate state,
which have exactly the same final state as the signal, and are the peaking
and non peaking backgrounds in the 2($\kk$) invariant mass
distribution. In addition, there are 43 background events from the decays
$\jp\to\phi f_1(1420)/f_1(1285)$ with $f_1$ decay to $\kk\pi^{0}$ and
$\jp\to\phi K^{*}(892)^{\pm}K^{\mp}$ with $K^{*}(892)^\pm$ decay to
$K^\pm\pi^0$, which have a final state of $\pi^02(\kk)$ similar to
that of the signal. These background decay channels have low detection
efficiency ($<0.1\%$), and do not produce a peak in the $\eta_c$
signal range. The expected yields of background events are 26 and 75 for the
peaking and non peaking backgrounds, respectively, determined with MC
simulation. As a cross-check, the backgrounds are also estimated with
the events in the $\phi$ sidebands region in data, and then using the MC information of the $\etac\to\phi\kk$ and $2(\kk)$ to scale the $\etac$ events in boxes $B,~C$ and $D$ to the signal region $A$, and total 104
events are obtained.

 To determine the $\eta_c\to\phi\phi$ yield, an amplitude analysis
 is performed on the selected candidate 1,276 events. We assume the
 observed candidates are from the process $J/\psi\to\gamma \phi\phi$
 with or without the $\eta_c$ intermediate state in the $\phi\phi$
 system. The amplitude formulas are constructed with the
 helicity-covariant method~\cite{chung2}, and shown in the appendix. The $\etac$ resonance is
 parametrized with the Breit-Wigner function multiplied by a damping factor
\begin{equation}
f(s)={1\over M^2-s-iM\Gamma}{\mathcal{F}(E_\gamma)\over \mathcal{F}(E^0_\gamma)},
\end{equation}
where $s$ is the square of $\phi\phi$ invariant mass, and $M$ and
$\Gamma$ are the $\etac$ mass and width, respectively. The damping factor is taken as
$\mathcal{F}(E_\gamma)=\exp(-{E_\gamma^2\over 16\beta^2})$ with $\beta=0.065$ GeV \cite{rhyan}, and the photon energy $E^0_\gamma$ corresponds to the $\sqrt{s}=M$.

In the analysis, the decay $J/\psi\to\gamma\eta_c\to\gamma\phi\phi$
and the nonresonant decays $J/\psi\to\gamma\phi\phi$ with different
quantum numbers $J^P$ (spin parity) in the $\phi\phi$ system are taken
into consideration. The differential cross section $d\sigma/d\Omega$
is calculated with
\begin{equation}
\begin{aligned}
{d\sigma\over d\Omega}=&\sum_{\textrm{helicities}}|A_{\etac}(\lambda_0,\lambda_\gamma,\lambda_1,\lambda_2)\\
&+\sum_{J^P} A_{NR}^{J^P}(\lambda_0,\lambda_\gamma,\lambda_1,\lambda_2) |^2,
\end{aligned}
\end{equation}
where $A_{\etac}$ is the amplitude for the
$\jp(\lambda_0)\to\gamma(\lambda_\gamma)\etac\to\gamma\phi(\lambda_1)\phi(\lambda_2)$,
with the joint helicity angle $\Omega$, and $A_{NR}^{J^P}$ is the
amplitude for the nonresonant decay $\jp\to\gamma\ff$ with $J^P$ for
the $\ff$ system. To simplify the fit, only the nonresonant components with $J^P=0^+,~0^-$ and $2^+$ are included, and the
components with higher spin are ignored. The symmetry of the identical
particles for the $\ff$-meson pair is implemented in the amplitude.

The magnitudes and phases of the coupling constants are determined
with an unbinned maximum likelihood fit to the selected
candidates. The likelihood function for observing the $N$ events in
the data sample is
\begin{equation}
\mathcal{L}=\prod_{i=1}^{N}P(x_i),
\end{equation}
where $P(x_i)$ is the probability to observe event $i$ with four
momenta $x_i=(p_\gamma,p_\phi,p_\phi)_i$, which is the normalized
differential cross section taking into account the detection
efficiency ($\epsilon_i$), and calculated by

\begin{equation}
P(x_i)={(d\sigma/d\Omega)_i \epsilon_i\over \sigma_{MC}},
\end{equation}
where the normalization factor $\sigma_{MC}$ can be calculated by a
signal MC sample $J/\psi\to\gamma\ff$ with $N_{MC}$ accepted
events. These events are generated with a phase space model and then
subjected to the detector simulation, and passed through the same
events selection criteria as applied to the data. With a MC sample
which is sufficiently large, $\sigma_{MC}$ is evaluated with
\begin{equation}
\sigma_{MC}={1\over N_{MC}}\sum_{i=1}^{N_{MC}}\left({d\sigma\over d\Omega}\right)_i.
\end{equation}

For a given N events data sample, the product of $\epsilon_i$ in
Eq.(3) is constant, and can be neglected in the fit. Rather than
maximizing $\mathcal{L}$, $\mathcal{T}=-\ln\mathcal{L}$ is minimized
using {\sc minuit}~\cite{minuit}.

In the analysis, the background contribution to the log-likelihood
value $(\ln\mathcal{L}_\textrm{bkg})$ is subtracted from the
log-likelihood value of data $(\ln\mathcal{L}_\textrm{data})$, i.e.
$\ln\mathcal{L} =
\ln\mathcal{L}_\textrm{data}-\ln\mathcal{L}_\textrm{bkg}$, where
$\ln\mathcal{L}_\textrm{bkg}$ is estimated with the MC simulated
background events, normalized to 101 events including peaking and
nonpeaking $\eta_c$ background.

In the fit, the mass and width of $\eta_c$ are fixed to the previous
BESIII measurements~\cite{etacBESIII}, i.e. $M=2.984$ GeV/$c^2$ and
$\Gamma=0.032$ GeV. The mass resolution of the $\etac$ is not
considered in the nominal fit, and its effect is considered as a
systematic uncertainty. The fit results are shown in
Fig.~\ref{fig:ffFitResult}, where the rightmost peak is due to
backgrounds from $\jp\to \phi\kk$ decay. The $\etac$ yield from the
fit is $N_{\etac}=549\pm65$, which is derived from numerical
integration of the resultant amplitudes, and the
statistical error is derived from the covariance matrix obtained from the fit.

To determine the goodness of fit, a global $\chi^2_{\textrm{g}}$ is
calculated by comparing data and fit projection histograms, defined as
\begin{equation}
\chi^2_{\textrm{g}} = \sum_{j=1}^{5}\chi^2_j \textrm{,~with~}\chi^2_j=\sum_{i=1}^N{(N_{ji}^\textrm{DT}-N_{ji}^\textrm{Fit})^2\over N_{ji}^{DT}},
\end{equation}
where $N_{ji}^{\textrm{DT}}$ and $N_{ji}^{\textrm{Fit}}$ are the
numbers of events in the $i$th bin of the $i$th kinematic variable
distribution. If $N_{ji}^\textrm{DT}$ is sufficiently large, the
$\chi^2_{\textrm{g}}$ is expected to statistically follow the $\chi^2$
distribution function with the number of degrees of freedom (ndf),
which is the total number of bins in histograms minus the number of
free parameters in the fit. In a histogram, bins with
less than ten events are merged with the nearby bins. The individual
$\chi^2_j$ give a qualitative evaluation of the fit quality for each
kinematic variable, as described in the following.

\begin{figure*}[!t]
\centering
\includegraphics[height=5cm,width=8cm]{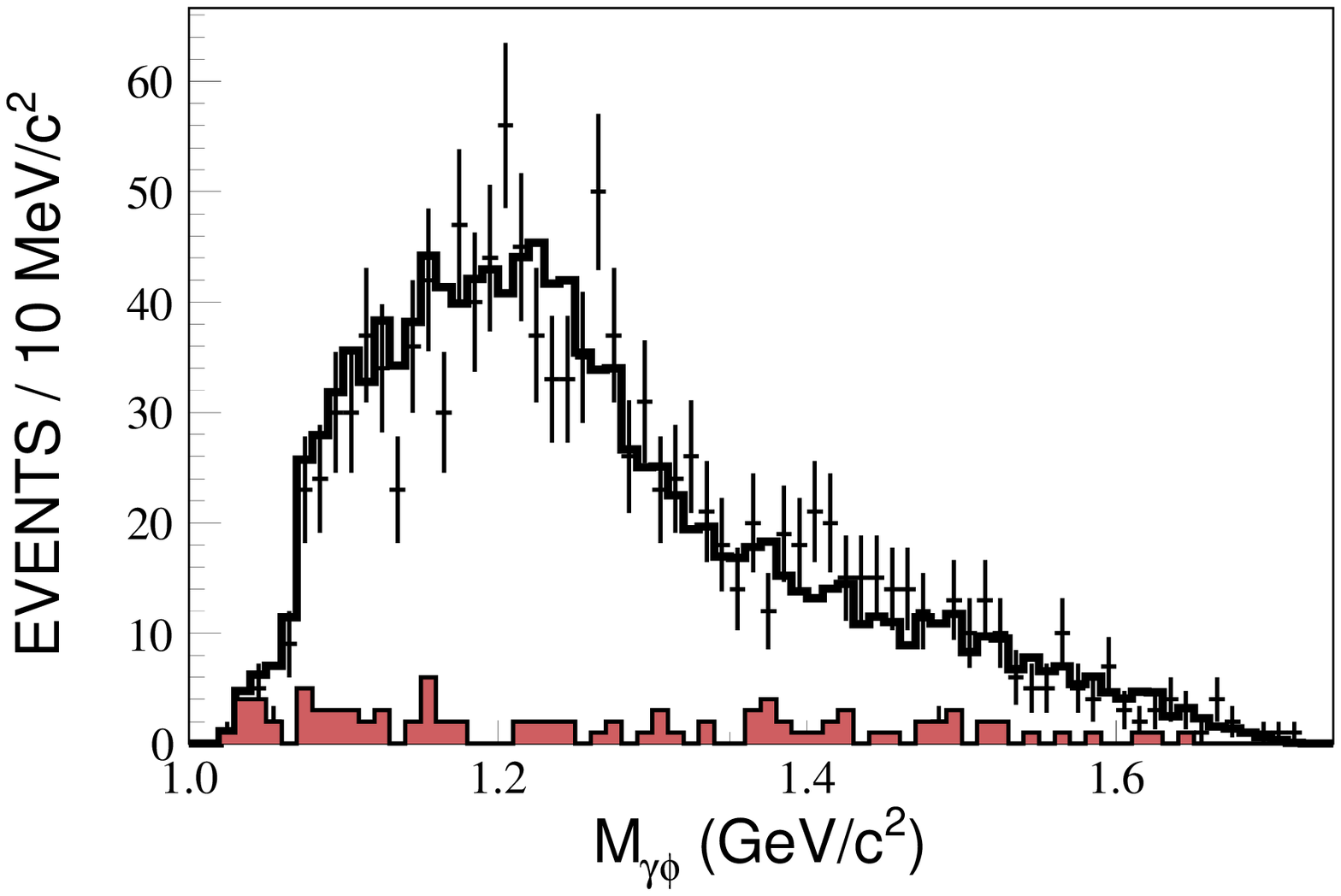}
\put(-25,125){\bf \large (a)}
\hspace{0.2cm}
\includegraphics[height=5cm,width=8cm]{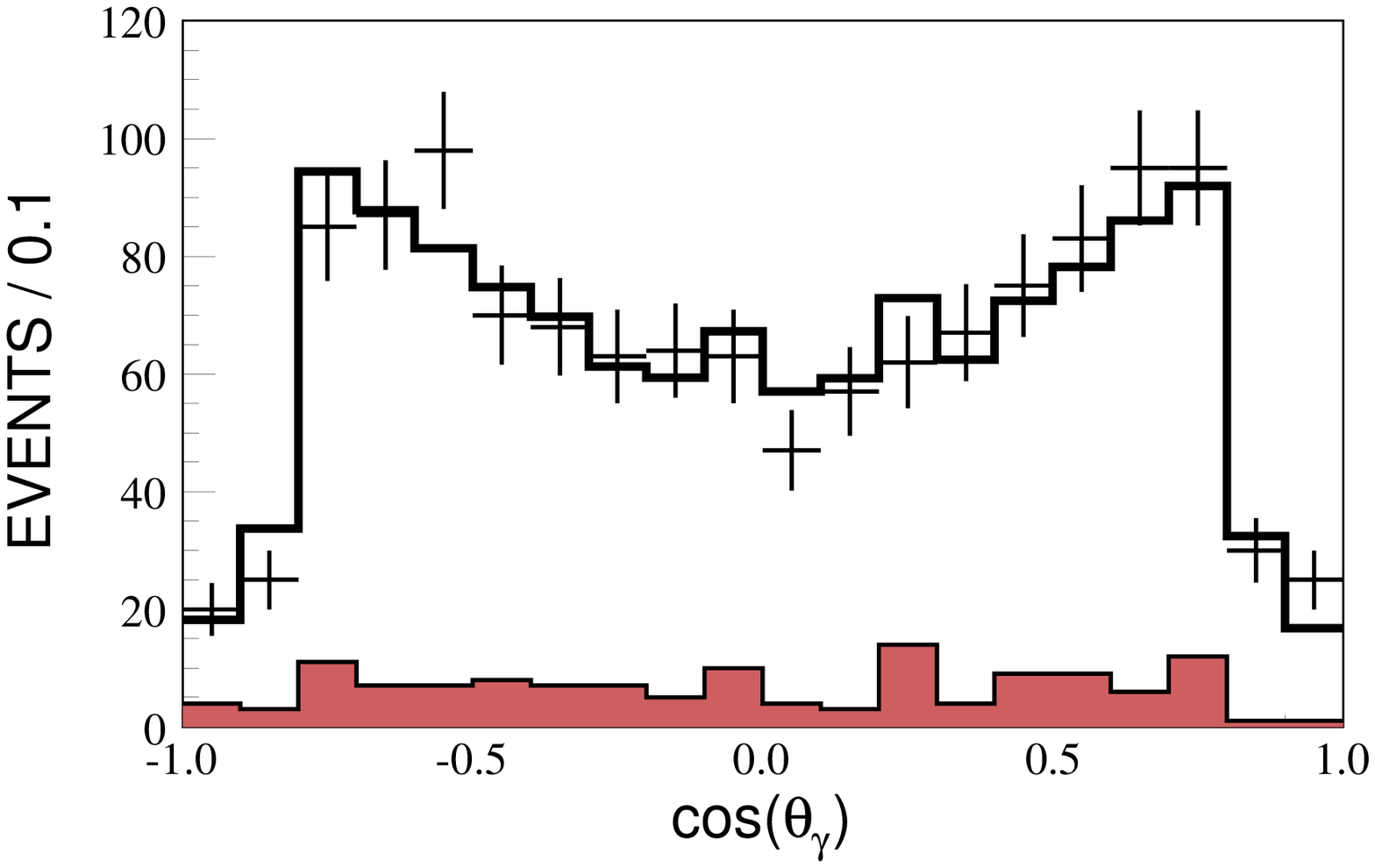}
\put(-30,125){\bf \large (b)}\\
\vspace{0.5cm}
\includegraphics[height=5cm,width=8cm]{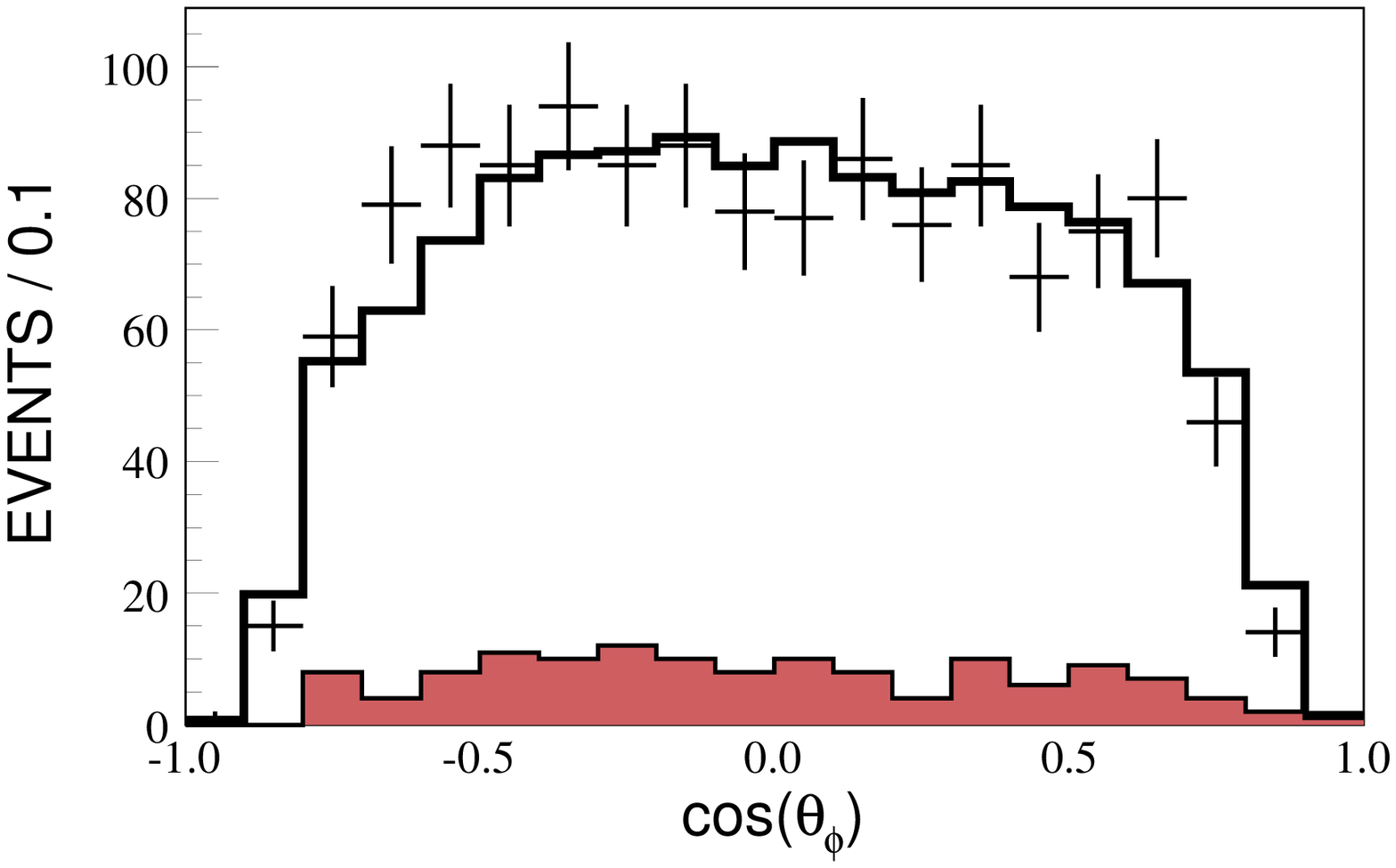}
\put(-28,125){\bf \large (c)}
\hspace{0.2cm}
\includegraphics[height=5cm,width=7.7cm]{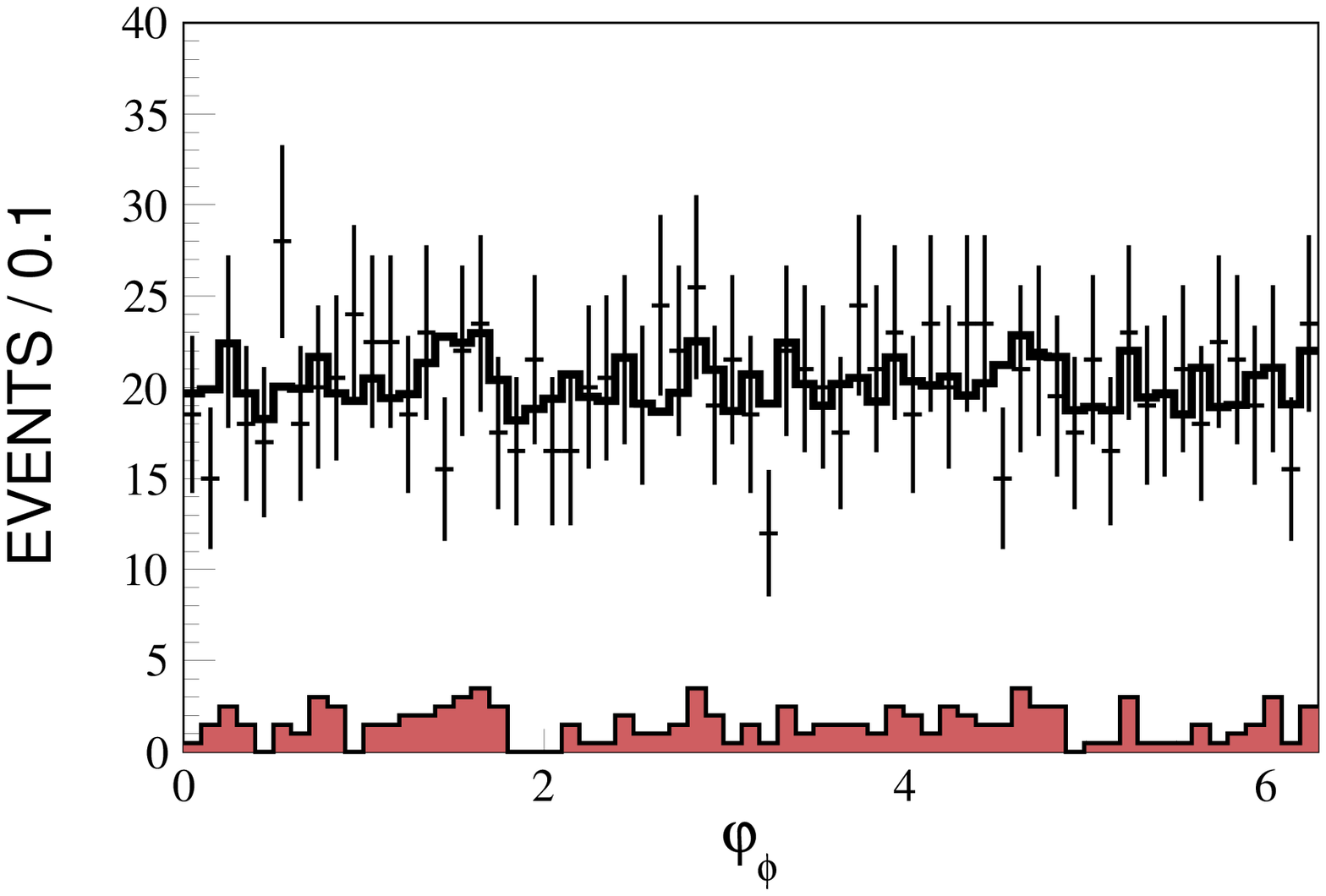}
\put(-25,125){\bf \large (d)} \renewcommand{\figurename}{Fig.}
{\caption{Distributions of (a) the $\gamma\phi$
    invariant mass M$_{\gamma\phi}$; (b) the polar angular of the
    photon $\cos\theta_\gamma$; (c) the polar angular of $\phi$ mesons
    $\cos\theta_\phi$; (d)and the azimuthal angular of $\phi$ mesons
    $\varphi_\phi$. The dots with error bar are the data, the solid
    line histograms represent the total fit results, and the filled
    histograms are the non-$J/\psi\to\gamma \phi \phi$ backgrounds
    estimated with the exclusive MC samples.}
\label{fig:goodness}}
\end{figure*}

Five independent variables are necessary to describe the three-body
decay $J/\psi\to\gamma\ff$. These are chosen to be the mass of the
$\ff$-system ($M_{\ff}$), the mass of the $\gamma \phi$-system
($M_{\gamma \phi}$), the polar angle of the $\gamma$
($\theta_{\gamma}$), the polar angle ($\theta_{\phi}$) and azimuthal
angle ($\varphi_{\phi}$) of the $\phi$-meson, where the angles are
defined in the $\jp$ rest frame. Figure~\ref{fig:goodness} shows the
comparison of the distributions of $M_{\gamma\phi}$ and angles between
the global fit and the data. A sum of all of the $\chi^2_j$ values gives
$\chi^2_{\text{g}}=215$ with ndf=191. The quality of the global fit
($\chi^2_{\textrm{g}}/$ndf) is 1.1, which indicates good agreement between
data and the fit results.

To validate the robustness of the fit procedure, a pseudodata sample
is generated with the amplitude model with all parameters fixed to the
fit results. A total of 2936 events are selected with the same
selection criteria as applied to the data. An identical fit process is
carried out, and the ratio of output $\etac$ signal yield to input
number of events is $1.03\pm0.03$.

\subsection{Search for $\etac\to\of$}
\begin{figure}[!t]
   \centering
  \includegraphics[width=0.9\columnwidth]{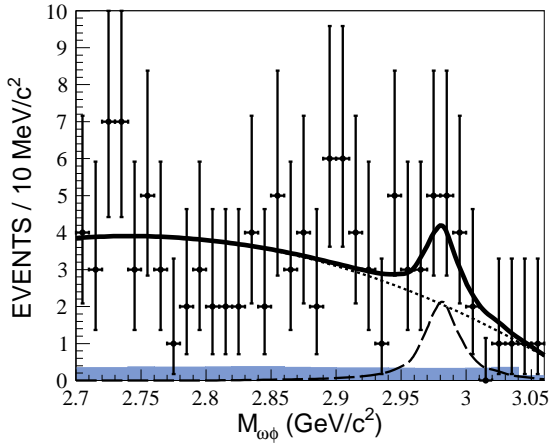}
   \renewcommand{\figurename}{Fig.}
   \caption{Results of the best fit to the $M_{\of}$
     distribution. Dots with error bars are data, the solid curve is
     the best fit result, corresponding to a $\eta_c$ signal yield of
     $10\pm6$ events, the shaded histogram is the background
     estimated from exclusive MC samples, the dashed curve indicates
     the $\etac$ signal, and the dotted curve is the fitted
     background.}
   \label{fig:omegaphimass}
 \end{figure}

Figure~\ref{fig:omegaphimass} shows the $\of$ invariant mass
distribution in the range from 2.70 to 3.05 GeV/$c^2$ for the
selected candidate events of $J/\psi\to\gamma\omega\phi$, and no
significant $\eta_c$ signal is observed. The background events from
$\jp$ decays are dominated by $\jp\to\eta'\phi$ with
$\eta'\to\gamma\omega$. A small amount of background is from the decays
$\jp\to f_0(980)\omega\to\kk\omega$ and $\jp\to f_X\omega\to
\pi^0\kk\omega$, where $f_X$ stands for the $f_1(1285)$ and
$f_1(1420)$ resonances. The sum of all above backgrounds estimated
from inclusive MC samples is small compared to the total number of
selected candidates and appears as a flat $M_{\of}$ distribution, as
shown in Fig.~\ref{fig:omegaphimass}.

 \begin{figure}[!t]
   \centering
  \includegraphics[width=0.9\columnwidth]{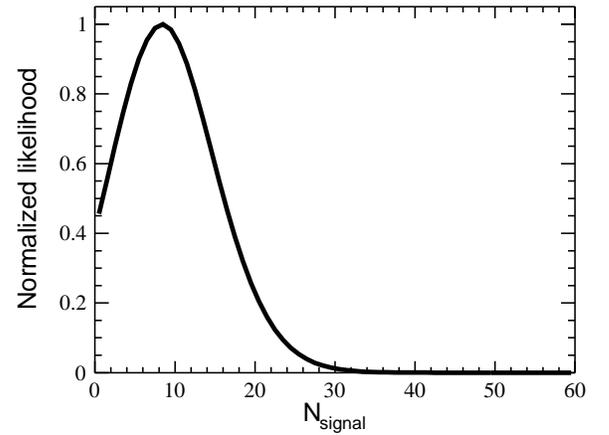}
   \renewcommand{\figurename}{Fig.}
   \caption{Normalized likelihood distribution versus the $\eta_c$
     yield for $\eta_c\to\of$.}
   \label{fig:omegaphiLikelihood}
 \end{figure}

 To set an upper limit for the branching fraction $Br(\etac\to\of$),
 the signal yield is calculated at the 90\% C.L. by a Bayesian
 method~\cite{pdg}, according to the distribution of normalized
 likelihood values versus signal yield, which is obtained from the
 fits by fixing the $\eta_c$ signal yield at different values.

 In the fit, the shape for the $\eta_c$ signal is described by the
 MC simulated line shape by setting the mass and width of $\eta_c$ to
 the BESIII measurement~\cite{etacBESIII}; the known background
 estimated with MC simulation is fixed in shape and magnitude in the
 fit; and the others are described by a second-order Chebychev
 function with floating parameters. The distribution of normalized
 likelihood values is shown in Fig. \ref{fig:omegaphiLikelihood}, and the
 upper limit of signal yield at the 90\% C.L. is calculated to be 18.

 To check the robustness of the event selection criteria, especially
 the dependence on $Br(\etac\to\of)$, the requirements of
 kinematic fit $\chi^2$ and $\phi/\omega$ mass windows are
 reoptimized with the measured upper limit. The $\etac$ signal yield
 is reestimated and is consistent within the statistical errors.

\section{Systematic uncertainties}
The following sources of systematic uncertainties are considered in the measurements of branching fractions.
\begin{enumerate}
\item Number of $\jp$ events\\
The number of $\jp$ events is determined using its hadronic
decays. The uncertainty is 0.6\%~\cite{jsinumber}.

\item{ Photon detection efficiency}\\
The soft and hard photon detection efficiencies are studied using the control samples $\psip\to\pi^0\pi^0\jp$, with $\jp$ decay $\ee$ or $\mu^+\mu^-$ and $\jp\to\rho\pi\to\pi^+\pi^-\pi^0$, respectively. The difference in the photon detection efficiency between the MC simulation and data is 1\%, which is taken as the systematic uncertainty.

\item Kaon/pion tracking and PID efficiency\\
  The uncertainties of kaon/pion tracking and PID efficiency are
  studied using the control samples $\jp\to\pp p\bar p$ and $\jp\to
  K^0_SK\pi$, with the decay $K_S^0\to\pi^-\pi^+$~\cite{tracking}. The
  uncertainties for tracking and PID efficiencies are both determined to be
  1\% per track.

\item Branching fractions\\
The uncertainties of branching fractions for $\jp\to\gamma\etac,~\phi\to\kk$, and $\omega\to\pp\pi^0$ are taken from the PDG~\cite{pdg}.

\item Kinematic fit\\
  To estimate the uncertainty associated with the $\chi^2$ requirement
  of the kinematic fit for the final state $\gamma2(\kk)$, we select
  the candidate events of $\jp\to\gamma\ff$ by requiring
  $\chi^2<20,~60$ or 150, and the $\etac$ signal yields are
  reestimated with amplitude analysis. The largest deviation to the
  nominal branching fraction, 6.7\%, is taken as the systematic
  uncertainty.

  For the final states $\gamma\kk\pp\pi^0$, we redetermine the upper
  limit on the branching fraction with the alternative requirement of
  the kinematic fit $\chi^2<20, 30, 50$ or 60, and the largest deviation
  to the nominal value, 2.4\% at $\chi^2<30$, is taken as the
  systematic uncertainty.

\item Mass window\\
  The uncertainties associated with the $\phi/\omega$ mass-window
  requirement arise if the mass resolution is not consistent between
  the data and MC simulation. The uncertainty related to the
  $\phi$-mass window requirement is determined with the control sample
  $\psip\to\gamma\chi_{cJ},~\chi_{cJ}\to\ff,$ and $\phi\to\kk$. The
  difference in $\phi$-selection efficiency is estimated to be 0.7\%
  and 1.1\% for the $\etac\to\ff$ and $\etac\to\of$ modes,
  respectively, where the different uncertainties obtained for the two
  decay modes are due to the different mass-window requirements. The
  uncertainty related with the $\omega$ mass-window requirement is
  determined with the control sample $\jp\to\omega\eta$ with
  $\omega\to\ppp$ and $\eta\to\ppp$. The difference in $\omega$
  selection efficiency is estimated to be 1.5\% for the $\etac\to\of$
  mode.

\item Background\\
  In the analysis of $J/\psi\to\gamma\phi\phi$, the uncertainty
  associated with the peaking background from
  $\jp\to\gamma\etac,~\etac\to\phi\kk$, and 2$(\kk)$ as well as the
  other unknown background is estimated by varying up or down the
  numbers of background events by one standard deviation according
  to the uncertainties of branching fractions in PDG~\cite{pdg}. The
  largest change in the $\eta_c\to\phi\phi$ signal yield is determined
  to be 0.9\%, and is taken as the systematic uncertainty.

  In the study of $J/\psi\to\gamma\omega\phi$, the uncertainty
  associated with the unknown background is estimated by replacing the
  second-order Chebychev function with the first-order one. The change
  of the upper limit of signal events is negligible. The uncertainty
  associated with the dominant background,
  $\jp\to\eta'\phi\to\gamma\of$, is estimated by varying the branching
  fraction by one standard deviation when normalizing the background
  in the fit. The difference in the resulting upper limit is
  determined to be 5.6\%, and is taken as the systematic uncertainty.

\item Fit range\\
  In the nominal fit, the fit range is set to be M$_{\phi\phi}$ and
  M$_{\omega\phi} >$ 2.70 GeV/c$^2$. Its uncertainty is estimated by
  setting the range of $M_{\ff}$ and $M_{\of}>2.60,~2.65,~2.75$ or 2.80
  GeV$/c^2$. The branching fraction of $\etac\to\ff$ and
  the upper limit for $\etac\to\of$ are reestimated.  The largest
  deviations to the nominal results, 0.7\% for the decay
  $\eta_c\to\phi\phi$ and 0.2\% for the decay $\eta_c\to\omega\phi$,
  are taken as the systematic uncertainties.

\item $\etac$ mass and width\\
Uncertainties associated with the $\eta_c$ mass and width are estimated by the alternative fits with the PDG values for the $\eta_c$ parameters~\cite{pdg}. The resulting differences in the $\eta_c$ signal yield, 1.3\% for $\eta_c\to\phi \phi$ and 5.6\% for $\eta_c\to\omega \phi$, are taken as systematic uncertainties.

\item Amplitude analysis\\
  Systematic uncertainties associated with the amplitude analysis
  arise including the uncertainties of the non-$\eta_c$ component and
  the mass resolution of $\eta_c$.

  In the nominal fit, the non-$\etac$ component is described by the
  nonresonant $\ff$-system assigned with quantum number $J^P=0^-,~0^+$
  and $2^+$. The statistical significance for the component with
  different $J^P$ is determined according to the difference of
  log-likelihood value between the cases with and without this
  component included in the fit, taking into account the change in the
  number of degrees of freedom. The significances for the non-$\eta_c$
  component with $J^P=0^-,~2^+,~0^+$ are 2.8$\sigma$, 3.0$\sigma$ and
  0.1$\sigma$, respectively. If the $0^-$ component is removed, the uncertainty is estimated to be +6.7\%.
  If the $2^+$ component is removed, the uncertainty is estimated to be $-26.0$\% mainly due to the strong interference between the $\eta_c$ and the $0^-$ components.

  The uncertainty related with the $\eta_c$ mass resolution is
  estimated by the alternative amplitude analysis with the detected
  width of the $\eta_c$ set to 34.2 MeV, estimated from the MC
  simulation with the nominal input $\eta_c$ width 32.0 MeV from
  Ref.~\cite{etacBESIII}. The resulting difference of the $\eta_c$ signal
  yield with respect to the nominal value is 2.2\%.

The total uncertainty from the amplitude analysis is estimated to be $^{+7.1\%}_{-26.1\%}$.
\end{enumerate}

Table~\ref{tab:syserror} summarizes all sources of systematic
uncertainties. The combined uncertainty is the quadratic sum of all
uncertainties except for that associated with
$Br(\jp\to\gamma\eta_c)$.

\begin{table}
\setlength{\tabcolsep}{0.5pc}
\caption{Summary of all systematic uncertainties from the different resources (\%). The combined uncertainty excludes the uncertainty associated with $Br(\jp\to\gamma\etac)$, which is given separately.\label{tab:syserror}}
\begin{tabular}{lcc}
\hline\hline
Sources     & $\etac\to\ff$ & $\etac\to\of$\\\hline
$N_{\jp}$   & 0.6    &0.6 \\
Photon      & 1.0    &3.0 \\
Tracking    & 4.0    &4.0 \\
PID         & ---     &4.0 \\
$Br(\phi\to\kk)$        & 2.0   & 1.0 \\
$Br(\omega\to\ppp)$ &--- &0.8 \\
Kinematic fit     & 6.7  & 2.4 \\
$M_{\kk}$ mass   & 0.7  & 1.1 \\
$M_{\ppp}$ mass  &---    & 1.5 \\
Background  & 0.9  & 5.6 \\
Fit range &0.7     &0.2     \\
$\etac$ mass and width& 1.3 & 5.6\\
Amplitude analysis & $^{+7.1}_{-26.1}$ & ---\\
Combined  & $^{+11.0}_{-27.4}$    & 10.7 \\\hline
$Br(\jp\to\gamma\etac)$ & 23.5  & 23.5 \\
\hline\hline
\end{tabular}
\end{table}

\section{Branching fractions}
\subsection{$\etac\to\ff$}
The product branching fraction of $\jp\to\gamma\eta_c\to\gamma\ff$ is calculated by
\begin{eqnarray*}
Br(\jp\to\gamma\eta_c)Br(\eta_c\to\ff)={N_\textrm{sig} \over N_{\jp}\epsilon Br^2(\phi\to\kk)}\\
=({4.3\pm0.5}(\textrm{stat})^{+0.5}_{-1.2}(\textrm{syst}))\times10^{-5},
\end{eqnarray*}
where $Br(\phi\to\kk)$ is the branching fraction of the $\phi\to\kk$ decay
taken from the PDG~\cite{pdg}, $N_\textrm{sig}$ is the $\etac$ signal
yield, and $\epsilon=24\%$ is the detection efficiency, determined
with the MC sample generated with the amplitude model with parameters
fixed according to the fit results. The number of $J/\psi$ events is
$N_{\jp}=223.7\times 10^6$~\cite{jsinumber}.

Using $Br(\jp\to\gamma\eta_c)=(1.7\pm0.4)\%$~\cite{pdg}, $Br(\etac\to\ff)$ is calculated to be
\begin{eqnarray*}
Br(\eta_c\to\ff)=\qquad\qquad\qquad\qquad\qquad\qquad\qquad\\
(2.5\pm0.3(\textrm{stat})^{+0.3}_{-0.7}(\textrm{syst})\pm0.6(\textrm{Br}))\times10^{-3},
\end{eqnarray*}
where the third uncertainty, which is dominant, is from the
uncertainty of $Br(\jp\to\gamma\etac)$, and the second uncertainty is
the quadratic sum of all other systematic uncertainties.

\subsection{$\etac\to\of$}
No significant signal is observed for $\etac\to\omega\phi$, and we determine
the upper limit at the 90\% C.L. for its branching fraction,
\begin{equation}
\begin{aligned}
Br(\etac\to\omega\phi)&<{N_\textrm{up} \over N_{\jp}\epsilon Br (1-\sigma_\textrm{sys})}\\&=2.5\times10^{-4},
\end{aligned}
\end{equation}
where $N_\textrm{up}=18$ is the upper limit on the number of $\etac$
events at the 90\% C.L., $\epsilon=5.9\%$ is the detection efficiency,
$\sigma_\textrm{sys}=25.8\%$ is the total systematic error, and $Br$
is the product branching fractions for the decay
$J/\psi\to\gamma\eta_c$, $\phi\to\kk$ and $\omega\to\ppp$~\cite{pdg}.

\section{SUMMARY AND DISCUSSION}
Using 223.7 million $\jp$ events accumulated with the BESIII detector,
we perform an improved measurement on the decay of $\etac\to\ff$. The
measured branching fraction is listed in Table~\ref{tab:br}, and
compared with the previous measurements. Within one standard
deviation, our result is consistent with the previous measurements,
but the precision is improved. No significant signal for $\etac\to\of$
is observed. The upper limit at the 90\% C.L. on the branching
fraction is determined to be $Br(\eta_c\to\of)<2.5\times10^{-4}$,
which is 1 order in magnitude more stringent than the previous upper limit~\cite{pdg}.

\begin{table*}[thbp]
  \caption{Comparison of BESIII measured $Br(\etac\to\ff)$ with the previous results and theoretical predictions, where the branching fractions of $\eta_c\to\phi\phi$ from BESII and DM2 are recalculated with Br$(\jp\to\gamma\etac)=(1.7\pm0.4)\%$~\cite{pdg}. \label{tab:br}}
\centering
\begin{tabular}{ccc}
\hline\hline
Experiment &$Br(\jp\to\gamma\etac)Br(\etac\to\ff)(\times10^{-5})$&$Br(\etac\to\ff)~(\times10^{-3})$  \\ \hline
BESIII     &$4.3\pm0.5^{+0.5}_{-1.2}$&$2.5\pm0.3^{+0.3}_{-0.7}\pm0.6$ \\
BESII~\cite{bes2etac2pp}&$3.3\pm0.8$&$1.9\pm0.6$ \\
DM2~\cite{dm2}        &$3.9\pm1.1$&$2.3\pm0.8$\\\hline
Theoretical &Prediction &$Br(\etac\to\ff)$ ($\times10^{-3}$)\\\hline
&pQCD\cite{pqcd} &$(0.7\sim0.8)$\\
&$^3P_0$ quark model~\cite{tpz}&$(1.9\sim2.0)$\\
&Charm meson loop~\cite{mesonloop}&2.0\\
\hline\hline
\end{tabular}
\end{table*}

The measured branching fractions of $\etac\to\ff$ are three times larger
than that calculated by next-to-leading perturbative QCD (pQCD) together with higher
twist contributions~\cite{pqcd}. This discrepancy between data and the
HSR expectation~\cite{hsr} implies that nonperturbative mechanisms play an important role in charmonium decay. To understand
the HSR violation mechanism, a comparison between the experimental
measurements and the theoretical predictions based on the light quark
mass correction~\cite{quarkmasscorrection}, the $^3P_0$ quark pair
creation mechanism~\cite{tpz} and the intermediate meson loop
effects~\cite{mesonloop} is presented in Table \ref{tab:br}. We note
that the measured $Br(\etac\to\ff)$ is close to the predictions of the
$^3P_0$ quark model~\cite{tpz} and the meson loop
effects~\cite{mesonloop}. In addition, the measured upper limit for
$Br(\etac\to\of)$ is comparable with the predicted value
$3.25\times10^{-4}$ in Ref.~\cite{mesonloop}. The consistency between
data and the theoretical calculation indicates the importance of QCD
higher twist contributions or the presence of a non-pQCD
mechanism.

\section{ACKNOWLEDGMENT}
The BESIII collaboration thanks the staff of BEPCII and the IHEP computing center for their strong support. This work is supported in part by National Key Basic Research Program of China under Contract No. 2015CB856700; National Natural Science Foundation of China (NSFC) under Contracts No. 11505034, No. 11375205, No. 11565006, No. 11647309, No. 11125525, No. 11235011, No. 11322544, No. 11335008, No. 11425524, and No. 11305090; the Chinese Academy of Sciences (CAS) Large-Scale Scientific Facility Program; the CAS Center for Excellence in Particle Physics (CCEPP); the Collaborative Innovation Center for Particles and Interactions (CICPI); Joint Large-Scale Scientific Facility Funds of the NSFC and CAS under Contracts No. 11179007, No. U1232201, and No. U1332201; CAS under Contracts No. KJCX2-YW-N29, and No. KJCX2-YW-N45; 100 Talents Program of CAS; National 1000 Talents Program of China; INPAC and Shanghai Key Laboratory for Particle Physics and Cosmology; German Research Foundation DFG under Collaborative Research Contract No. CRC 1044, and No. FOR 2359; Istituto Nazionale di Fisica Nucleare, Italy; Koninklijke Nederlandse Akademie van Wetenschappen (KNAW) under Contract No. 530-4CDP03; Ministry of Development of Turkey under Contract No. DPT2006K-120470; Russian Foundation for Basic Research under Contract No. 14-07-91152; the Swedish Resarch Council; U. S. Department of Energy under Contracts No. DE-FG02-05ER41374, No. DE-SC-0010504, No. DE-SC0012069, and No. DESC0010118; U.S. National Science Foundation; University of Groningen (RuG) and the Helmholtzzentrum fuer Schwerionenforschung GmbH (GSI), Darmstadt; and WCU Program of National Research Foundation of Korea under Contract No. R32-2008-000-10155-0.

\newpage
\begin{center}{\bf\huge Appendix} \\ {\bf\large Amplitude analysis of the decays $J/\psi\to\gamma\phi\phi$}\end{center}
\begin{appendices}
\section{Amplitudes\label{appAmplitude}}
For the decay
$\jp(\lambda_0)\to\gamma(\lambda_\gamma)\etac\to\gamma\phi(\lambda_1)\phi(\lambda_2)$, where the $\lambda_i(i=\gamma,0,1,2)$ indicates helicity values for the corresponding particles, the helicity-coupling amplitude is given by
\begin{eqnarray}\label{chungs}
A_{\etac}(\lambda_0,\lambda_\gamma,\lambda_1,\lambda_2)&=& F^\psi_{\lambda_\gamma}(r_1)D^{1*}_{\lambda_0,-\lambda_\gamma}(\theta_0,\phi_0)BW_j(m_{\ff})\nonumber\\
&\times&F^{\etac}_{\lambda_1,\lambda_2}(r_2)D^{0*}_{0,\lambda_1-\lambda_2}(\theta_{1},\phi_{1}){\mathcal{F}(E_\gamma)\over\mathcal{F}(E^0_\gamma)},\nonumber\\  \end{eqnarray}
where  $r_1$($r_2$) is the momentum
difference between $\gamma$ and $\eta_c$ (two $\phi$ mesons) in the rest frame of $\jp~(\eta_c)$, and $\theta_{0}~
(\phi_{0})$ and $\theta_{1}~(\phi_{1})$
are the polar (azimuthal) angles of the momentum vectors
${\bf P}_{\eta_c}$ and ${\bf~P}_\phi$ in the helicity system of $\jp$ and $\eta_c$, respectively. The $z$-axis defined for $\eta_c\to\phi(\lambda_1)\phi(\lambda_2)$ is taken along the outgoing direction of $\phi(\lambda_1)$ in the $\eta_c$ rest frame, and the $x$-axis is in the ${\bf P}_{\eta_c}$ and ${\bf~P}_{\phi(\lambda_1)}$ plane, which together with the new $y$-axis forms a right-hand system.
$BW_j(m)$ denotes the Breit-Wigner parametrization for the $\eta_c$ peak. The damping factor $\mathcal{F}(E_\gamma)$ is taken as $\mathcal{F}(E_\gamma)=\exp(-{E_\gamma^2\over 16\beta^2})$ with $\beta$=0.065 GeV \cite{rhyan}; $E_\gamma^0$ is the photon energy corresponding to $m_{\ff}=m_{\etac}$.
The helicity-coupling amplitudes $F^\psi_{\lambda_\gamma}$ and $F^{\eta_c}_{\lambda_1,\lambda_2}$ are related to the covariant amplitudes in the $LS$-coupling scheme by \cite{chung2}
\begin{eqnarray}\label{zerominus}
F^\psi_1&=&-F^\psi_{-1}={g_{11}\over \sqrt 2}r_1{B_1(r_1)\over B_1(r^0_1)},\nonumber\\
F^{\etac}_{1,1}&=&-F^{\etac}_{-1,-1}={g'_{11}\over \sqrt 2}r_2{B_1(r_2)\over B_1(r^0_2) },\nonumber\\
F^{\etac}_{0,0}&=&0,
\end{eqnarray}
where $B_l(r)$ is the Blatt-Weisskopf factor \cite{chung2}, $r_1^0$ and $r_2^0$ indicate the momentum differences for the two decays with $m_{\ff}=m_{\eta_c}$, and $g_{ls}$ and $g'_{ls}$ are the coupling constants for the two decays.

For the direct decay $\jp\to\gamma\ff$, the mass spectrum of $\ff$ appears as a smooth distribution within the $\etac$ signal region; hence the Breit-Wigner function is excluded. The amplitudes for the direct decay are decomposed into partial waves associated with the $\ff$-system with  quantum numbers $J^P=0^-,~ 0^+$ and $2^+$, and the high spin waves are neglected. These amplitudes are taken as

\begin{eqnarray}\label{chungs}
A^{0^-}_{NR}(\lambda_0,\lambda_\gamma,\lambda_1,\lambda_2)&=& F^{\psi}_{\lambda_\gamma,0}D^{1*}_{\lambda_0,-\lambda_\gamma}(\theta_0,\phi_0)F^{0^-}_{\lambda_1,\lambda_2}\nonumber\\
&\times&D^{0*}_{0,\lambda_1-\lambda_2}(\theta_{1},\phi_{1})\nonumber \textrm{~for $0^-$},\\
A^{0^+}_{NR}(\lambda_0,\lambda_\gamma,\lambda_1,\lambda_2)&=& F^{\psi}_{\lambda_\gamma,0}D^{1*}_{\lambda_0,-\lambda_\gamma}(\theta_0,\phi_0)F^{0^+}_{\lambda_1,\lambda_2}\nonumber\\
&\times&D^{0*}_{0,\lambda_1-\lambda_2}(\theta_{1},\phi_{1})\nonumber \textrm{~for $0^+$},\\
A^{2^+}_{NR}(\lambda_0,-\lambda_\gamma,\lambda_1,\lambda_2)&=&\sum_{\lambda_J} F^{\psi}_{\lambda_\gamma,\lambda_J}D^{1*}_{\lambda_0,\lambda_J-\lambda_\gamma}(\theta_0,\phi_0)\nonumber\\
&\times&F^{2^+}_{\lambda_1,\lambda_2}D^{2*}_{\lambda_J,\lambda_1-\lambda_2}(\theta_{1},\phi_{1})\nonumber \textrm{~for $2^+$}.
\end{eqnarray}
Here, helicity-coupling amplitudes $F_{\lambda_1,\lambda_2}^{J^P}$ are related to covariant amplitudes.
For $J^P=0^-$,  helicity amplitudes take the same form as that in Eq. (\ref{zerominus}).

For the $0^+$ case, helicity amplitudes are taken as
\begin{eqnarray}
F_1^{\psi}&=&F^{\psi}_{-1}={g_{21}r_1^2 \over \sqrt{6}}{B_2(r_1)\over B_2(r^0_1)}+{g_{01}\over \sqrt3},\nonumber\\
F^{0^+}_{11}&=&F^{0^+}_{11}={g'_{22}r^{2}_2\over \sqrt6}{B_2(r_2)\over B_2(r^0_2)} + {g'_{00}\over \sqrt3},\\
F^{0^+}_{00}&=&\sqrt{2\over 3}r_2^{2}g'_{22}{B_2(r_2)\over B_2(r^0_2)}- {g'_{00}\over \sqrt3}.\nonumber
\end{eqnarray}

For the $2^+$ case, helicity amplitudes are taken as
\begin{eqnarray}
F_{12}^{\psi}&=&F^{\psi}_{-1-2}={g_{43}r_1^4 \over \sqrt{70}}{B_4(r_1)\over B_4(r^0_1)}+{g_{21}r_1^2\over \sqrt{10}}{B_2(r_1)\over B_2(r^0_1)}\nonumber\\
&-&{g_{22}r_1^2\over \sqrt6}{B_2(r_1)\over B_2(r^0_1)}+\sqrt{2\over 105}g_{23}r_1^2{B_2(r_1)\over B_2(r^0_1)}+{g_{01}\over \sqrt5},\nonumber\\
\end{eqnarray}
\begin{eqnarray}
F_{11}^{\psi}&=&F^{\psi}_{-1-1}={-2g_{43}r_1^4 \over \sqrt{35}}{B_4(r_1)\over B_4(r^0_1)}-{g_{21}r_1^2\over \sqrt5}{B_2(r_1)\over B_2(r^0_1)}\nonumber\\
&+&\sqrt{3\over 35}g_{23}r_1^2{B_2(r_1)\over B_2(r^0_1)}+{g_{01}\over \sqrt{10}},\nonumber\\
\end{eqnarray}
\begin{eqnarray}
F_{10}^{\psi}&=&F^{\psi}_{-10}=\sqrt{3\over 35}g_{43}r_1^4{B_4(r_1)\over B_4(r^0_1)} +{g_{21}r_1^2\over 2\sqrt{15}}{B_2(r_1)\over B_2(r^0_1)}\nonumber\\
&+&{1\over 2}g_{22}r_1^2{B_2(r_1)\over B_2(r^0_1)}+{2g_{23}r_1^2\over \sqrt{35}}{B_2(r_1)\over B_2(r^0_1)}+{g_{01}\over \sqrt{30}},\nonumber\\
\end{eqnarray}
\begin{eqnarray}
F^{2^+}_{11}&=&F^{2^+}_{-1-1}=\sqrt{3\over 35}g'_{42}{B_4(r)\over B_4(r')}r^4+{g'_{20}r^{2}_2\over \sqrt3}{B_2(r_2)\over B_2(r^0_2)}\nonumber\\
&-&{g'_{22}r_2^2\over \sqrt{21}}{B_2(r_2)\over B_2(r^0_2)} + {g'_{02}\over \sqrt{30}},\nonumber\\
\end{eqnarray}
\begin{eqnarray}
F^{2^+}_{10}&=&F^{2^+}_{-10}=-{2\over \sqrt{35}}g'_{42}r^4{B_4(r)\over B_4(r')}-{1\over 2}g'_{21}r_2^2{B_2(r_2)\over B_2(r^0_2)}\nonumber\\
&-&{g'_{22}r_2^{2}\over 2\sqrt7}{B_2(r_2)\over B_2(r^0_2)}+{g'_{02}\over \sqrt{10}},\nonumber\\
\end{eqnarray}
\begin{eqnarray}
F^{2^+}_{1-1}&=&F^{2^+}_{-11}={g_{42}r^4\over \sqrt{70}}{B_4(r)\over B_4(r')}+\sqrt{2\over 7}g'_{22}r_2^2{B_2(r_2)\over B_2(r^0_2)}\nonumber\\
&+&{g'_{02}\over \sqrt{5}}.
\end{eqnarray}
For these nonresonant decays, the differences of momenta $r_l^0$ are calculated at the value $m_{\ff}=2.55$ GeV.

The total amplitude is expressed by:
\begin{eqnarray}\label{}
A(\lambda_0,\lambda_\gamma,\lambda_1,\lambda_2)&=&A_{\etac}(\lambda_0,\lambda_\gamma,\lambda_1,\lambda_2)
\nonumber\\
&+&\sum_{J^P}A^{J^P}_{NR}(\lambda_0,\lambda_\gamma,\lambda_1,\lambda_2),
\end{eqnarray}
where the sum runs over $J^P=0^-,0^+$ and $2^+$, and the symmetry of identical particle for two $\phi$ mesons is implied by exchanging their helicities and momentum.
The differential cross section is given by
\begin{eqnarray}\label{}
d\Gamma&=&\left(3\over
8\pi^2\right)\sum_{\lambda_0,\lambda_\gamma,                    \lambda_1,\lambda_2}A(\lambda_0,\lambda_\gamma,\lambda_1,\lambda_2)\nonumber\\
&\times&A^*(\lambda_0,\lambda_\gamma,\lambda_1,\lambda_2)d\Phi,
\end{eqnarray}
where $\lambda_0,\lambda_\gamma=\pm1$, and $\lambda_1,\lambda_2=\pm1,0$, and
$d\Phi$ is the element of standard three-body phase space.

\section{Fit method}
The relative magnitudes and phases for coupling constants are determined
by an unbinned maximum likelihood fit. The joint probability density
for observing $N$ events in the data sample is
\begin{equation}
\mathcal{L}=\prod_{i=1}^N P(x_i),
\end{equation}
where $P(x_i)$ is a probability to produce event $i$ with a set of four-vector momentum $x_i=(p_{\gamma},p_{\phi},p_{\phi})_i$. The normalized $P(x_i)$ is calculated from the differential cross section
\begin{equation}
P(x_i)={(d\Gamma /d\Phi)_i \over \sigma_{MC}},
\end{equation}
where the normalization factor $\sigma_{MC}$ is calculated from a MC
sample with $N_{MC}$ accepted events, which are generated with a phase
space model and then subject to the detector simulation, and are
passed through the same event selection criteria as applied to the
data analysis. With a MC sample of sufficiently large size, the
$\sigma_{MC}$ is evaluated with
\begin{equation}
\sigma_{MC}={1\over N_{MC}}\sum_{i=1}^{N_{MC}}\left({d\Gamma\over d\Phi}\right)_i.
\end{equation}
For technical reasons, rather than maximizing $\mathcal{L}$, $S=-\ln\mathcal{L}$ is minimized using the package MINUIT. To subtract the background events, the $\ln\mathcal{L}$ function is replaced with

\begin{equation}
\ln\mathcal{L} = \ln\mathcal{L}_\textrm{data}-\ln\mathcal{L}_\textrm{bg}.
\end{equation}

After the parameters are determined in the fit, the signal yields of a given resonance can be estimated by scaling its cross section ratio $R_i$ to the number of net events, {\it i.e.},
\begin{equation}
N_i=R_i*(N_\textrm{obs}-N_\textrm{bg}),\textrm{~with~}R_i={\Gamma_i\over \Gamma_\textrm{tot}},
\end{equation}
where $\Gamma_i$ is the cross section for the $i$th resonance, $\Gamma_\textrm{tot}$ is the total cross section, and $N_\textrm{obs}$ and $N_\textrm{bg}$ are the numbers of observed events and background events, respectively.

The statistical error, $\delta N_i$, associated with signal yields $N_i$ is estimated based on the covariance matrix, $V$, obtained from the fit according to
\begin{equation}
\delta N_{i}^{2} = \sum_{m=1}^{N_\textrm{pars}}\sum_{n=1}^{N_\textrm{pars}}\left({\partial N_i\over \partial X_m}{\partial N_i\over \partial X_n}\right)_{\bf{X}={\bf \mu}}V_{mn}({\bf X}),
\end{equation}
where ${\bf X}$ is a vector containing parameters, and ${\bf \mu}$ contains the fitted values for all parameters. The sum runs over all $N_\textrm{pars}$ parameters.

\section{Results of parameters}
The nominal fit includes the decays, $\jp\to\gamma\eta_c\to\gamma\ff$ and $\jp\to\gamma(\ff)_{J^P}\to\gamma\ff$ with $J^P=0^-,0^+,2^+$. The coupling constants $g_{ls}$ are taken as complex numbers, and they are recombined to give new reduced parameters, which are determined in the fit. The reduced parameters are listed in Table \ref{parlist}, and the fitted values are given in Table \ref{fitvalues}.

\begin{table}
\caption{Definition of reduced parameters for decays in the nominal fit.\label{parlist}}
\begin{tabular}{cl}
\hline\hline
Decays & Reduced parameters \\\hline
$\jp\to\gamma\etac\to\gamma\ff$& $N_0=g_{11}g'_{11}$\\
$\jp\to\gamma(\ff)_{0^-}\to\gamma\ff$& $N_1=g_{11}g'_{11}$\\
$\jp\to\gamma(\ff)_{2^+}\to\gamma\ff$& $N_2=g_{43}g'_{22},\tilde{g}_{21}=g_{21}/g_{43}$,\\
& $\tilde{g}_{22}=g_{22}/g_{43},\tilde{g}_{23}=g_{23}/g_{43}$,\\
&$\tilde{g}_{20}=g_{20}/g_{43},\tilde{g}'_{20}=g'_{20}/g'_{22}$,\\
&$\tilde{g}'_{21}=g'_{21}/g'_{22},\tilde{g}'_{02}=g'_{02}/g'_{22}$,\\
&$\tilde{g}'_{42}=g'_{42}/g'_{22}$\\
$\jp\to\gamma(\ff)_{0^+}\to\gamma\ff$& $N_3=g_{01}g'_{00},\tilde{h}_{21}=g_{21}/g_{01}$,\\
&$\tilde{h}'_{22}=g'_{22}/g'_{00}$.\\
\hline\hline
\end{tabular}
\end{table}

\begin{table}
\caption{The values of reduced parameters determined in the nominal fit.\label{fitvalues}}
\begin{tabular}{ccc}
\hline\hline
Parameter $z$ & Magnitude $|z|$ & Argument arg$(z)$ /($2\pi$)\\\hline
N0 & $0.11\pm0.01$   &  $0.65\pm0.05$\\
N1 & $0.12\pm0.01$  & $0.13\pm0.05$\\
N2 & $0.59\pm0.27$  & $0.87\pm0.07$\\
$\tilde{g}_{21}$   & 0.29$\pm$0.12 & 0.59$\pm$0.07\\
$\tilde{g}_{22}$   & 0.36$\pm$0.14 & 0.90$\pm$0.07\\
$\tilde{g}_{23}$   & 0.43$\pm$ 0.31& 0.96$\pm$0.11\\
$\tilde{g}_{20}$   & 0.07$\pm$0.04 & 0.54$\pm$0.10\\
$\tilde{g}'_{20}$  & 1.00$\pm$0.54 &0.61$\pm$0.09\\
$\tilde{g}'_{21}$  & 0.00$\pm$0.51 &0.21$\pm$0.34\\
$\tilde{g}'_{02}$  & 0.66$\pm$ 0.28&0.50$\pm$0.08\\
$\tilde{g}'_{42}$  & 0.59$\pm$0.26  &0.50$\pm$0.08\\
N3 & $0.01\pm0.00$  & $0.52\pm0.07$\\
$\tilde{h}'_{21}$  & 1.00$\pm$0.90  &0.99$\pm$0.98\\
$\tilde{h}'_{22}$  & 2.35$\pm$1.25  &0.89$\pm$0.09\\\hline\hline
\end{tabular}
\end{table}

\end{appendices}

\newpage

\end{document}